# Fission Parameters Measurements for Np, Pu, Am, and Cm Isotopes inside a Salt Blanket Micromodel

Yury E. Titarenko, Oleg V. Shvedov, Vyacheslav N. Konev, Mikhail M. Igumnov,
Vyacheslav F. Batyaev, Evgeny I. Karpikhin, Valery M. Zhivun, Aleksander B. Koldobsky, Ruslan D. Mulambetov, Dmitry V. Fischenko, Svetlana V. Kvasova

*Institute for Theoretical and Experimental Physics, B.Cheremushkinskaya 25,117259 Moscow, Russia[1].*

Eduard F. Fomushkin, Vyacheslav V. Gavrilov, Gleb F. Novoselov

*Russian Federal Nuclear Center - VNIIEF, 607190, Sarov (Arzamas-16),Nizhny Novgorod region, Russia*

Aleksander V. Lopatkin, Viktor G. Muratov

*Research and Development Institute of Power Engineering,P.O.B. 788, Moscow 101000, Russia*

Aleksander F. Lositskiy, Boris L. Kurushin

*JSC Chepetsky Mechanical Plant, Belov st. 7, 427600 Glazov,Udmurt Republic, Russia*

Stepan G. Mashnik

*Los Alamos National Laboratory, Los Alamos, NM 87545, USA*

Hideshi Yasuda

*Japan Atomic Energy Research Institute, Tokai, Ibaraki, 319-1195, Japan*

**Abstract -***Pursuing verification of the nuclear data for actinides, we have made a run of experiments to determine fission reaction rates in facilities with different neutron spectra. The researches of the kind are particularly argent when going over from the transmutation physics studies to designing the transmutation reactors and developing their fuel cycle equipment. In this case, the nuclear data on the minor actinides (Np, Am, Cm) are notably interesting with the view to correct prediction of transmutation rates and to validation of hazardous nuclear and radiation environment for the external (off-reactor) fuel cycle. It is in the case of just those nuclides when the well-known ENDF/B6 and JENDL3.2 libraries give the most discrepant nuclear cross sections, thus necessitating the high-priority experimental tests.*

*The MAKET zero-power heavy water reactor has been used at ITEP to measure the fission characteristics of the Np, Pu, Am, and Cm isotopes in the $0.52NaF+0.48ZrF_4$ melt-filled salt blanket micromodel.*

*The $^{237}Np(n,f)$, $^{238}Pu(n,f)$, $^{239}Pu(n,f)$, $^{240}Pu(n,f)$, $^{241}Pu(n,f)$, $^{242m}Am(n,f)$, $^{243}Cm(n,f)$, $^{245}Cm(n,f)$, $^{247}Cm(n,f)$, $^{238}U(n,f)$, $^{238}U(n,\gamma)$, $^{235}U(n,f)$ fission reaction rates have been measured.*

*The neutron spectrum in the isotope irradiation locations was monitored by measuring the rates of the ($^{235}U(n,f)$, $^{238}U(n,\gamma)$, $^{55}Mn(n,\gamma)$, $^{63}Cu(n,\gamma)$, $^{176}Lu(n,\gamma)$, $^{197}Au(n,\gamma)$, $^{115}In(n,n')$, $^{27}Al(n,\alpha)$, and $^{64}Zn(n,p)$) reactions whose cross sections have been commonly accepted.*

*The measured functionals are compared with the respective results of MCNP code simulation obtained using the ENDF/B6 and JENDL3.2 neutron databases.*

---

[1]*E-mail: Yury.Titarenko@itep.ru*

## I. INTRODUCTION

We are of the opinion that the accuracy requirements for the knowledge of the neutron cross sections of the minor actinides (MA) are determined by the scope of the tasks to be tackled, namely,

- knowledge of the spectrum-averaged cross sections of Np to 10%, Am to 15-20%, and Cm to 30-50% is quite sufficient when studying the radiation balance of nuclear power production and in predicting the radiation property variations in the fuel irradiated in the PWRs;
- a 5-10% accurate MA cross sections seem to be proper when designing nuclear reactors with side MA transmutation, i.e., small MA addends (1-3% of the basic fuel mass) to the reactor core;
- the MA cross sections must be known to at least the same accuracy as the U (0.2-0.3\%) and Pu (0.5%) isotope cross sections when studying and designing specialized reactors-transmuters (molten salt reactors, accelerator-driven systems) with a large MA portion of the fuel.

The true accuracy can be estimated in the integral experiments by comparing between the experimental and calculated fission functionals in terms of different databases. It is the experiments with the salt system-emulating micromodels that are discussed in the present work.

## II. DESCRIPTION OF THE MAKET ZERO-POWER FACILITY

The ITEP salt blanket micromodel (SBM) is cylinder-shaped of 230-mm diameter and 522-mm height. The SBM was fastened on the central fuel channel (FC) at a 400-mm height above the lower fuel lattice in the core of the MAKET zero-power facility. The entire SBM channel structure (see Fig. 1) has been made of zirconium at Chepetsky Mechanical Plant (Glazov, Russia). The SBM is filled with the $0.52NaF + 0.48ZrF_4$ melt. An additional salt insert, which raises the salt melt content, or bushing-type fuel elements can be placed in the central FC of a 58-mm external diameter. The experimental samples were irradiated on being placed in the $0.52NaF + 0.48ZrF_4$ melt-filled containers, which were downed into the channels located at 46.5-mm, 72.0-mm, and 96.5-mm radial distances from the SBM axis (0 mm is the center of the additional salt insert).

The experiments were supported by forming two 100-mm step hexagonal fuel lattices assembled of FCs with the bushing-type fuel elements that contained uranium of 90% $^{235}$U enrichment. One of the fuel lattices (21-1) was assembled of 34 FCs and 67,272 g of the melt (the salt insert in the central FC). Another lattice (21-2) was assembled of 33 FCs and 65,592 g of the melt (the fuel elements in the central FC).

For the two lattices, the MAKET facility absolute power value, which corresponds to the rated neutron flux density $\varphi_0$, was determined by the techniques described in [1]. The results are presented in Table 1.

Table I
The experimental values of the MAKET power (wt).

|  | 21-1 | | 21-2 | |
|---|---|---|---|---|
|  | W ± ΔW | % | W ± ΔW | % |
| $W_0$ | 81.4 ± 1.7 | 2.1 | 78.8 ± 1.6 | 2.0 |

The constant monitoring of the neutron flux density for either of the lattices has made it possible to determine the absolute power value of the MAKET facility as

$$W_i = W_0 \cdot \frac{\varphi_i}{\varphi_0}.$$

## III. REACTION RATE MEASUREMENT TECHNIQUES

The $^{235}$U, $^{237}$Np, $^{238}$Pu, $^{239}$Pu, $^{240}$Pu, $^{241}$Pu, $^{242m}$Am, $^{243}$Cm, $^{245}$Cm, and $^{247}$Cm fission reaction rates were measured by the solid-state nuclear track detector (SSNTD) techniques.

The high-enriched U, Pu, Am, and Cm samples were prepared in the SM-2 electromagnetic mass separator with the sectored magnetic field, $H=H_0R_0/R$, of mean radius $R_0=1000$ mm and a 2-radian ($114^0$) bending angle. The mass dispersion is 20 mm per 1% of relative difference in masses at a 60-100-mm total trajectory length.

The enriched samples were mass-spectrometered to not worse than 0.1% (at a $\geq 10^{-2}$ relative content of the impurity isotope) and to 1-3% (at a $<10^{-4}$ impurity content). In most of the samples, the basic isotope enrichment was not worse than 98% (as of the sample manufacture date).

After chemical purification, the fissile substance was applied to a stainless steel substrate by electrolysis of nitrates from aqueous or alcohol solutions. The spot diameter was 6 mm in all the target layers. The distribution uniformity of the active substance over a spot was tested by self-radiography.

The samples were "weighed" (i.e., the numbers of fissile nuclides were determined in the layers) by α- and γ-spectrometry methods and on the basis of spontaneous and thermal-neutron-induced fission of nuclei. The semiconductor detectors with the standard electronics outfit were used in the α- and γ-spectrometers. As regards the spontaneous and thermal neutron-induced fission, the "weighing" was realized by the same techniques as in the main measurements, i.e., SSNTDs were used. The ~3.3% weighing error arises mainly from the uncertainty in the number of nuclei in the benchmark sources, as well as from the errors in the periods of the spontaneous and thermal neutron-induced fission.

The SSNTDs used in the measurements were made of silicate glass and polycarbonate film of molecular mass 90,000. The detectors are insensitive to the α-, β-, γ-, and n-emissions and, given appropriate conditions, provide for 100% efficiency in recording the fission fragments.

The glass or polymer plates were placed in parallel to, or coaxially with, a fissile isotope layer at 6 mm from the layer in the containers called hereafter the integral chambers. A purposed diaphragm confined a circle to a 6-mm diameter on the detector surface. The accuracy of all the dimensions in the measurement chambers was monitored using an instrumental microscope.

The polymer film was used in so-called shutter chambers, wherein a device like camera shutter was mounted. The shutter chamber differs from the integral chamber in that the former is doubled, i.e., two samples can be irradiated simultaneously. Besides, the samples and the film detectors in the shutter chamber are interlayered with metal plates with orifices (the shutters). During irradiations, the shutters move upwards, thus permitting the fission fragments to reach the detectors through the orifices. The shutters were set in motion by electromagnets actuated from the remote control board located in the MAKET facility control board room. The remote control of the shutters secured independent switching of the shutters in all four containers. In such a way, the irradiation time interval was fixed to a very high accuracy for each of the containers. This is especially important when operating with the samples of a high spontaneous fission activity, as well as in the case of neutron flux density variations when the irradiator operation mode is reached.

The geometric efficiency, $\Omega$, of the chamber (i.e., a probability for a fragment to appear in the detection domain) was calculated on assumption of a uniform distribution of a fissile substance over the layer. It should be noted that the $<\Omega N>$, value was measured directly when "weighing" the layers by spontaneous or thermal neutron-induced fission, with the non-uniformity of substance distribution over the layer being allowed for automatically.

In each of the measurement runs, two measurement chambers were placed into the SBM central channel, one with the experimental isotope layer and another with a $^{235}$U or $^{239}$Pu benchmark layer. Each of the experimental isotopes was irradiated in at least four runs, of which two irradiations were with the U benchmark, and two with the Pu benchmark.

The SSNTD residence time in a measurement chamber was strictly monitored with a view to making allowance for the background induced by spontaneous fission of nuclei. The residence time was found mostly to be 10-11 min.

The irradiations were made at atmospheric pressure (i.e., in disvacuumed chambers). The tentative study has shown that, given the selected geometric dimensions, the impact of atmospheric air on the fragment detection efficiency and, far less on the eventual measurement results, is negligible.

After irradiations, the glass detectors were take out of the chambers and, after that, etched chemically. An optic microscope was used to inspect the detectors and to count the tracks in each detector. Each of the detectors was inspected separately by two independent observers. As a rule, the differences between the readings did not exceed ~0.5%.

The algorithm for processing the relative measurement data is obvious enough. In processing, the contribution from the impurity isotopes to the total number of detected fission fragments was allowed for.

In this case, the fission reaction rate of each nuclide was determined as

$$R^X_{(n,f)} = \frac{T^X}{\xi^X \cdot N^X} \cdot \frac{1}{t} = \frac{T^X}{N^{Eff}} \cdot \frac{1}{t} \quad (1),$$

where $T^x$ is the number of the detected fission fragment tracks of the experimental nuclides; $N^x$ is the number of nuclei in the experimental "layers"; $\xi^x$ is the SSNTD sensitivity to a fission event ($\xi = \Omega \cdot \rho$, where is a probability for a fission fragment to be detected when it hits the detector surface,); $t$ is irradiation time.

The $^{235}$U(n,f), $^{238}$U(n,γ), $^{55}$Mn(n,γ), $^{63}$Cu(n,γ), $^{176}$Lu(n, γ), $^{197}$Au(n, γ), $^{115}$In(n,n'), $^{27}$Al(n,α), and $^{64}$Zn(n,p) reaction rates were measured by γ-spectrometry using a GC2518 Ge detector and the CANBERRA Co.-made electronics outfit. The spectrometer resolution was 1.8 keV in the 1332 keV γ-line. The reaction rates were determined as

$$R^{A\,B}_{x_0} = \frac{N^{A^*\,B}_{Z^*}}{N^{A\,B}_Z \cdot F(t_2, t_3)} =$$

$$= \left( \frac{SPA(t_3)}{\lambda \cdot \eta \cdot \varepsilon_{abs} \cdot F(t_2, t_3)} \right)^{A^*\,B}_{Z^*} \cdot \quad (2),$$

$$\cdot \left( \frac{M}{m \cdot x \cdot y} \right)^{A\,B}_Z \cdot \frac{1}{N_{Av}}$$

where $SPA(t_3)$ is the recorded γ-intensity induced by nuclide decay and reduced to the irradiation run end; $\lambda$ is the decay constant of product nuclide; $\eta$ is the absolute quantum yield for the respective energy of product nuclide; $\varepsilon_{abs}$ is the relative or absolute spectrometer efficiency for the measured γ-line energy; $M$ is molecular weight; $m$ is weight of an experimental sample; $x$ is the content of a given element in the sample; $y$ is the content of a given isotope in the element; $N_{Av}$ is the Avogadro number; $F(t_2,t_3)$ is the time function to allow for the nuclide decay during irradiation, considering the time correction for reaching the preset power of the MAKET critical facility.

The formula to calculate the $^{235}$U(n,f) reaction rate is the same as (2). The numbers of $^{235}$U nuclei in the experimental samples were determined by recording the 185.7 keV γ-line of $^{235}$U:

$$R_{(n,f)_0}^{^{235}U} = \frac{N^{f.p.}}{J^{f.p.} \cdot N^{^{235}U} \cdot F(t_2, t_3)} \quad (3),$$

where $N^{f.p.}$ is the number of fission product nuclei selected to be $^{143}$Ce and $^{97}$Zr; $J^{f.p.}$ is the $^{143}$Ce and $^{97}$Zr yields per a single fission event; $N^{^{235}U}$ is the number of $^{235}$U nuclei in an experimental sample.

The errors in the measured reaction rates were calculated by the standard error transfer formula. Figs. 2 and 3 show the overall results of determining the measured reaction rates relative to the $^{235}$U (n,f) reaction rate in the above mentioned two lattices. The measured $^{55}$Mn(n,γ), $^{63}$Cu(n,γ), $^{176}$Lu(n,γ), $^{197}$Au(n,γ), $^{115}$In(n,n'), and $^{27}$Al (n,α) reaction rates were used to restore the neutron spectra in the SBM channels. Fig. 4 shows the results of restoring the neutron spectra by the KASKAD code.

## IV SIMULATION OF REACTION RATES

The MCNP-4B code [2] was used to calculate the MAKET facility critical states and the reaction rates for the experimental samples placed in the SBM channels. The calculations were made in terms of two computational models of the facility (the input data files for the MCNP code), namely 21-1 for salt insert in the central fuel channel and 21-2 for fuel elements therein.

The models contain the detailed three-dimensional descriptions of the MAKET facility and SBM, including the description of two cylindrical tanks of the lower and upper supporting lattices. The fuel channels are of a complicated structure, which is a set of stacked ring-shaped fuel elements with fissile material. In the computational models, the fuel elements were taken to be a set of height-uniform cylindrical layers (zones).

The SBM description is absolutely in conformity with the respective drawings; all the actual dimensions, gaps, etc. have been included. The physical approximation to the description (i.e., a simplified presentation of the computational model element geometry that does not involve any strong distortion of the MAKET neutron- physics characteristics) was used solely for the SBM upper section above the heavy water level and for the small pieces (bolts, screws, etc.). In the actual experiments, the central axis of the assembled critical facility (with fixed SBM axis) did not coincide with the central axis of the tank. An additional computational analysis has shown that the shift of the reactor core to the center of the tank does not affect the calculation results. In the computational models, therefore, the SBM is placed at the heavy water tank center.

To facilitate the calculations we have compiled a library of neutron cross sections basing on the ENDF/B-VI [3] and JENDL-3.2 [4] estimated nuclear data files. All the neutron cross sections are presented to be continuous-energy data with a 0.1% accuracy of restoring the cross sections in the range of unresolved resonances. The library was compiled using the ML45 code generated by the authors of the present work on the basis of the NJOY-94(99) code [5,6]. The scattering matrices $S_{(\alpha,\beta)}$ in the thermal neutron range were constructed for deuterium and hydrogen to suit the temperature of heavy and light water in the experimental facility. The techniques for preparing the neutron cross sections have been validated by calculating the critical tests [7]. Thus, the computational models reproduced the true three-dimensional geometry of the MAKET facility with the SBM and described the nuclide composition of the structure element materials as plausibly as possible. The correctness of preparing the working models and the neutron cross section libraries has been tested by calculating the critical states of the MAKET facility with the SBM. The deviations from $k_{eff}$ =1 are $\Delta k_{eff}$ =0.037% for the first lattice and $\Delta k_{eff}$ = 0.431% for the second lattice. Each of the critical states was calculated using 10-15 million neutron histories.

The models having been tested and approved, the reaction rates were calculated for the samples placed in the respective containers. Basing on the tentative analysis and calculations, the true detector geometry and mass was assumed as described solely for $^{nat}$In and $^{238}$U to include the self-blocking effects. In all the rest samples, the reaction rates were simulated in terms of unified calculation in the homogeneous range that describes the locations of experimental samples and the averaged material composition of the containers. Each version was calculated using 50-60 million neutron histories. Figs. 3 and 4 present the simulation results too.

## V. COMPARISON BETWEEN CALCULATIONS AND EXPERIMENT

The comparison between calculations and experiment was made allowing for isotope composition of the samples.

In the comparison, use was made of a unified absolute normalization 1/(sWg). With that purpose, the measured reaction rates that had dimension [s$^{-1}$] were reduced to the form

$$R_{norm}^{\exp} = \frac{R_{x_0}^{^A_Z B} \cdot N_{Av}}{A \cdot W}$$

where W is the experimental MAKET facility absolute power corresponding to a given lattice;

$N_{Av}$ is the Avogadro number;

$R_{x_0}^{^A_Z B}$ is the measured reaction rate;

$A$ is the atomic weight of an element.

The simulated reaction rate values, which were initially normalized to the fission neutron number [1/n], have been renormalized as

$$R_{nopm}^{calc} = \frac{R_0}{E_f \cdot \Sigma_f} \cdot \frac{N_{Av}}{A}$$

where $E_f$ is the energy deposition in a single fission event, which is taken to be 194.0833 MeV, conforming to the MCNP-4B output;

$\Sigma_f$ is the total number of fission events in the MAKET facility after normalizing to a single fission neutron (the $\Sigma_f$ value is taken from the MCNP-4B output).

Figs. 2 and 3 show the comparison results. The mean square factor of the calculation-experiment deviations presented in Figs. 2 and 3 and in Table 2 was calculated as

$$<F> = 10^{\sqrt{<(\log(R_{cal,i}/R_{\exp,i}))^2>}}$$

where $< >$ designates averaging over all the experimental and calculated data. The presented data are indicative of the differences ranging from 2% to 47%[2] between the experiment and calculations for the reactions with well-known cross sections. In the case of the actinide reaction rates, the differences are ranging from 1% to 53%.

The comparison was made also by the conventional relative technique (i.e., by normalizing all the data to the $^{235}$U(n,f) reaction rate) to exclude the error in determining the MAKET facility power.

Figs. 4 and 5 show the experimental reaction rates renormalized to the $^{235}$U (n,f) reaction rate. The calculation-experiment deviations presented in Table 3 indicate the differences of 3-35%[1] between experiment and calculations in the relative normalization for the reactions with well-known cross sections. This means that the above two comparison methods give the identical experiment-calculation convergence, thereby confirming that the systematic errors are null in determining the MAKET facility power. The same follows from the identical experiment-calculation divergence level obtained by the two comparison methods for the actinide reaction rates.

To compare with the spectra restored by the KASKAD code, Fig. 5 displays the MCNP-4B code-calculated spectra.

VI. CONCLUSION

The accuracy requirements of the actinide fission cross sections have been stated in Introduction above. Tables 2 (the absolute normalization) and 3 (the relative normalization) present the experiment-calculation differences obtained using the ENDF/B-VI and JENDL-3.2 libraries for the thermal and intermediate spectra produced inside the SBM in different fuel lattices of the MAKET facility.

The tentative analysis (Table 4) of the experiment-calculation differences for two lattices and two normalizations has shown that

1. the JENDL-3.2 and ENDF/B-VI libraries show about the same preference when used for $^{235}$U, $^{238}$Pu, $^{239}$Pu, $^{240}$Pu, $^{241}$Pu, and $^{242m}$Am in both lattices;

2. JENDL-3.2 is more preferable for $^{237}$Np and $^{247}$Cm in both lattices;

3. ENDF/B-IV is more preferable for $^{238}$U in lattice 21-2; both libraries show the identical agreement for $^{238}$U in lattice 21-1-5(M2);

4. it depends on a particular (absolute or relative) normalization if either library is preferred for $^{243}$Cm and $^{245}$Cm. A better agreement with experiment is given by ENDF/B-VI for the absolute normalization and by JENDL-3.2 for the relative normalization.

Any conclusive inference concerning the advantages of ENDF/B-VI and JENDL-3.2 can only be drawn from analyzing the results obtained with the FKBM-2M and BIGR fast-spectrum facilities and from recertifying the $^{240}$Pu samples [8].

ACKNOLOEGMENT

The work has been performed under the ISTC Project #1145 was supported by Japan (JAERI). In part, the work has been supported by the U.S. Department of Energy.

---

[2] The $^{115}$In(n,2n)+$^{113}$In(n,γ) reaction sum was disregarded because the $^{113}$In(n,γ) reaction rate is much above the $^{115}$In(n,2n) reaction rate.

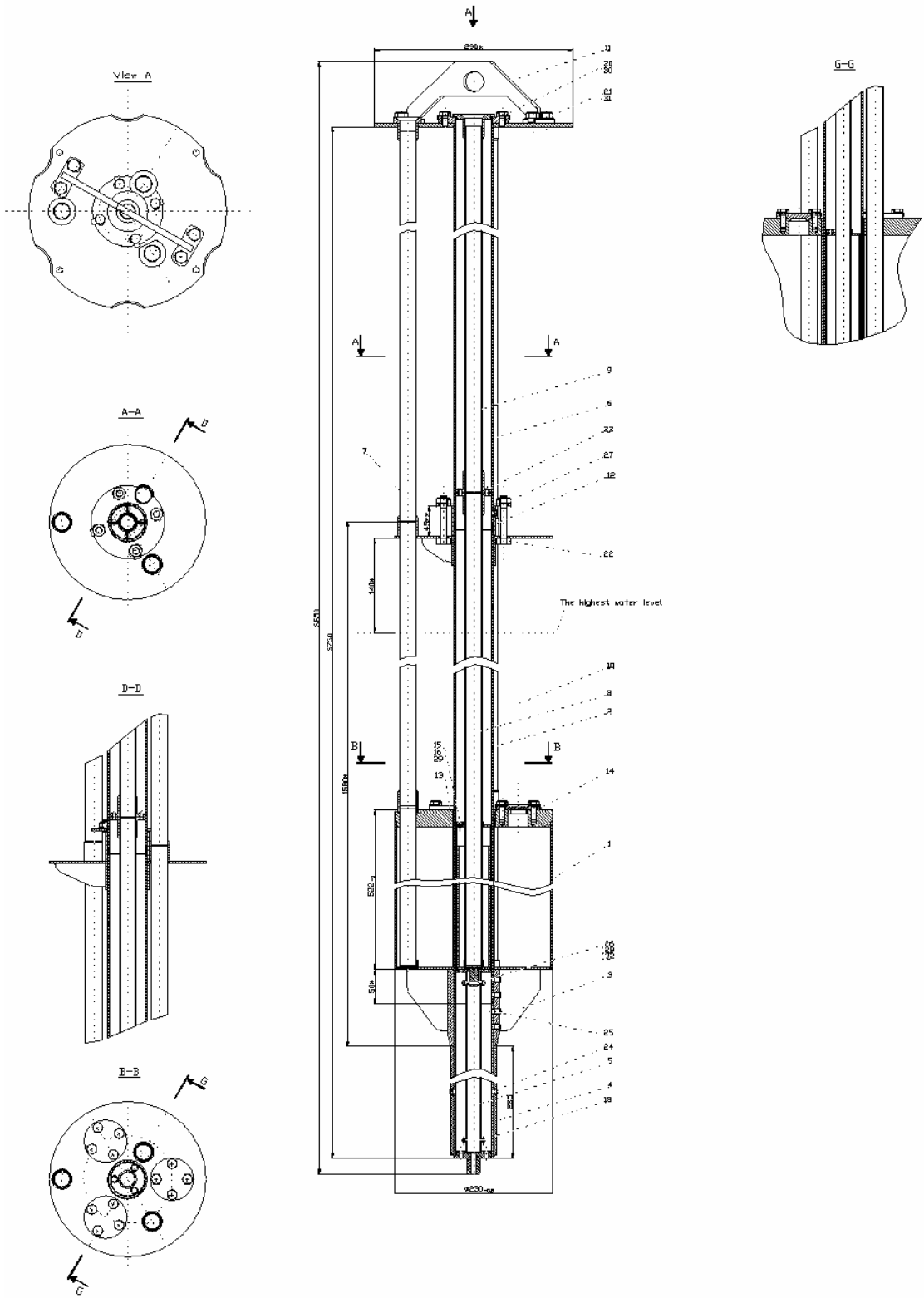

Fig. 1 . Salt blanket micromodel

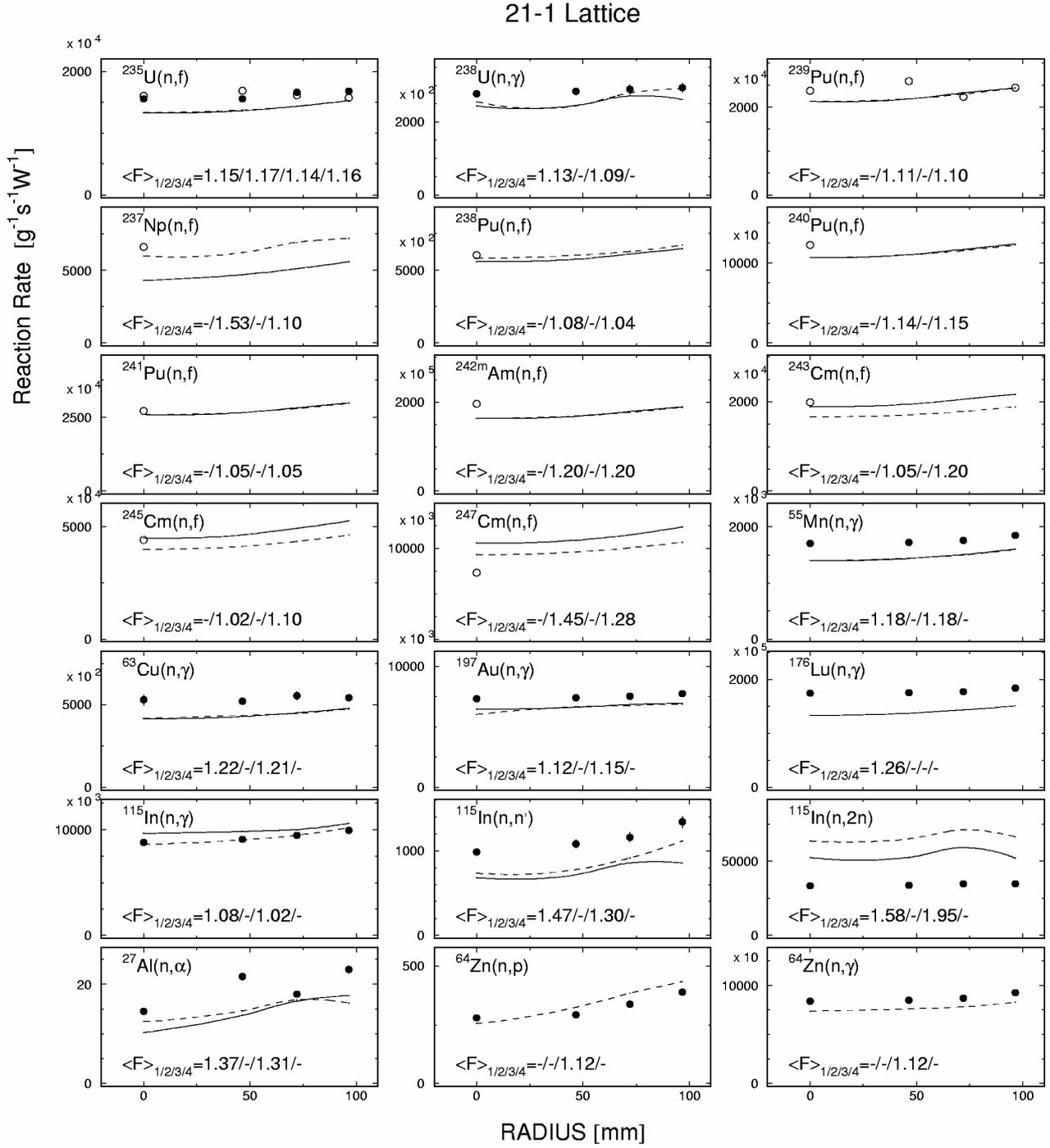

Fig. 2. The MAKET-measured and calculated reaction rates in the lattice with salt insert in central channel after normalizing to the $^{235}$U(n,f) reaction rate at the respective points. The black circles are the ITEP data. The light circles are the VNIIEF data. The dashed lines are the JENDL-3.2 –based calculations. The solid lines are the ENDF/B-IV–based calculations. $<F>_{1/2/3/4}$ are the experiment-calculation mean square deviation factors, where $<F>_1$ is the ITEP experiment- ENDF/B-VI calculation difference; $<F>_2$ is the VNIIEF experiment- ENDF/B-VI calculation difference; $<F>_3$ is the ITEP experiment- JENDL-3.2 calculation difference; $<F>_4$ is the VNIIEF experiment- JENDL-3.2 calculation difference. The gaps indicate lack of the appropriate experimental or calculated results.

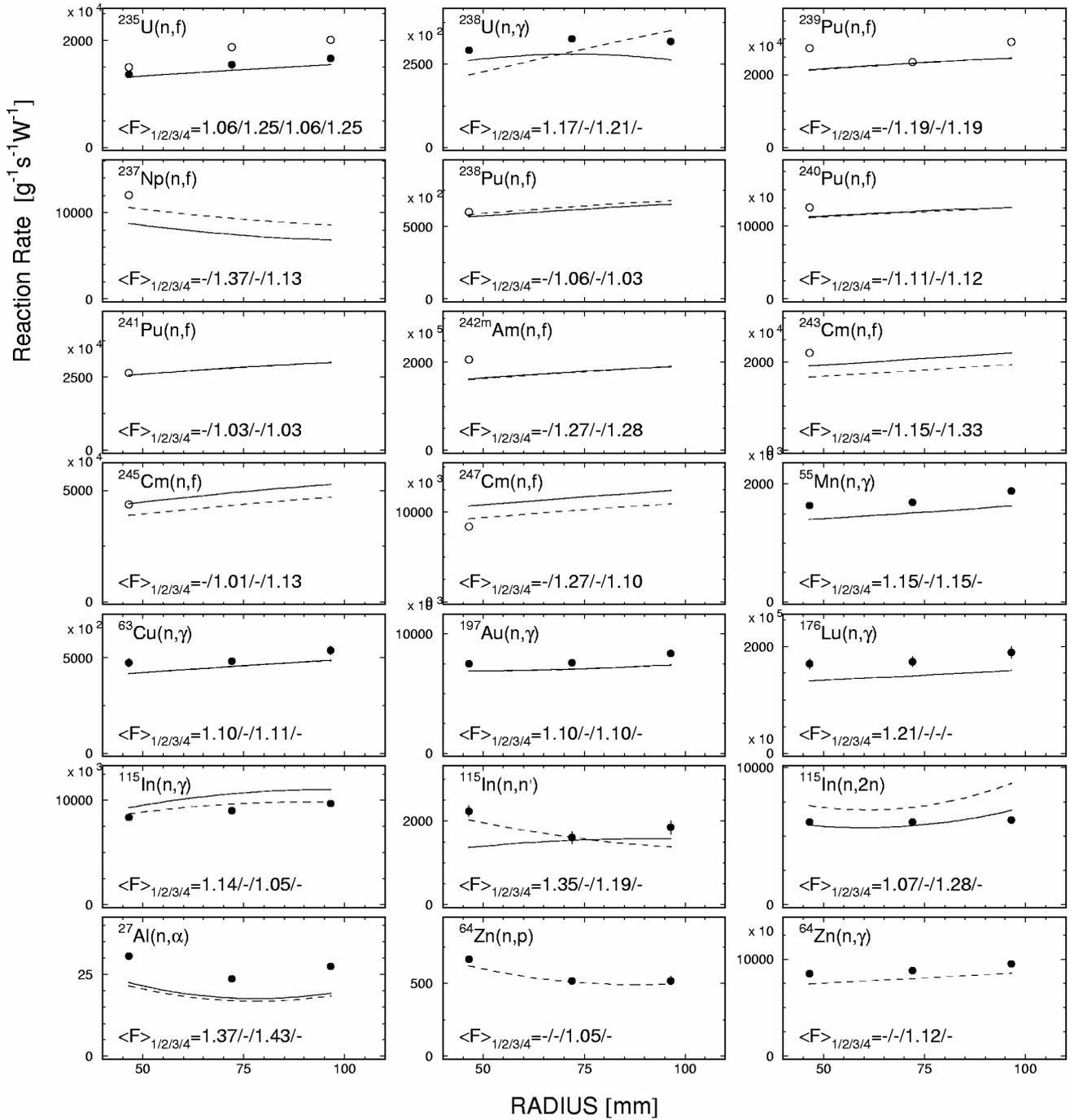

Fig. 3. The MAKET-measured and calculated reaction rates in the lattice with fuel elements in central channel after normalizing to the $^{235}$U(n,f) reaction rate at the respective points. The black circles are the ITEP data. The light circles are the VNIIEF data. The dashed lines are the JENDL-3.2 –based calculations. The solid lines are the ENDF/B-IV–based calculations. $<F>_{1/2/3/4}$ are the experiment-calculation mean square deviation factors, where $<F>_1$ is the ITEP experiment- ENDF/B-VI calculation difference; $<F>_2$ is the VNIIEF experiment- ENDF/B-VI calculation difference; $<F>_3$ is the ITEP experiment- JENDL-3.2 calculation difference; $<F>_4$ is the VNIIEF experiment- JENDL-3.2 calculation difference. The gaps indicate lack of the appropriate experimental or calculated results.

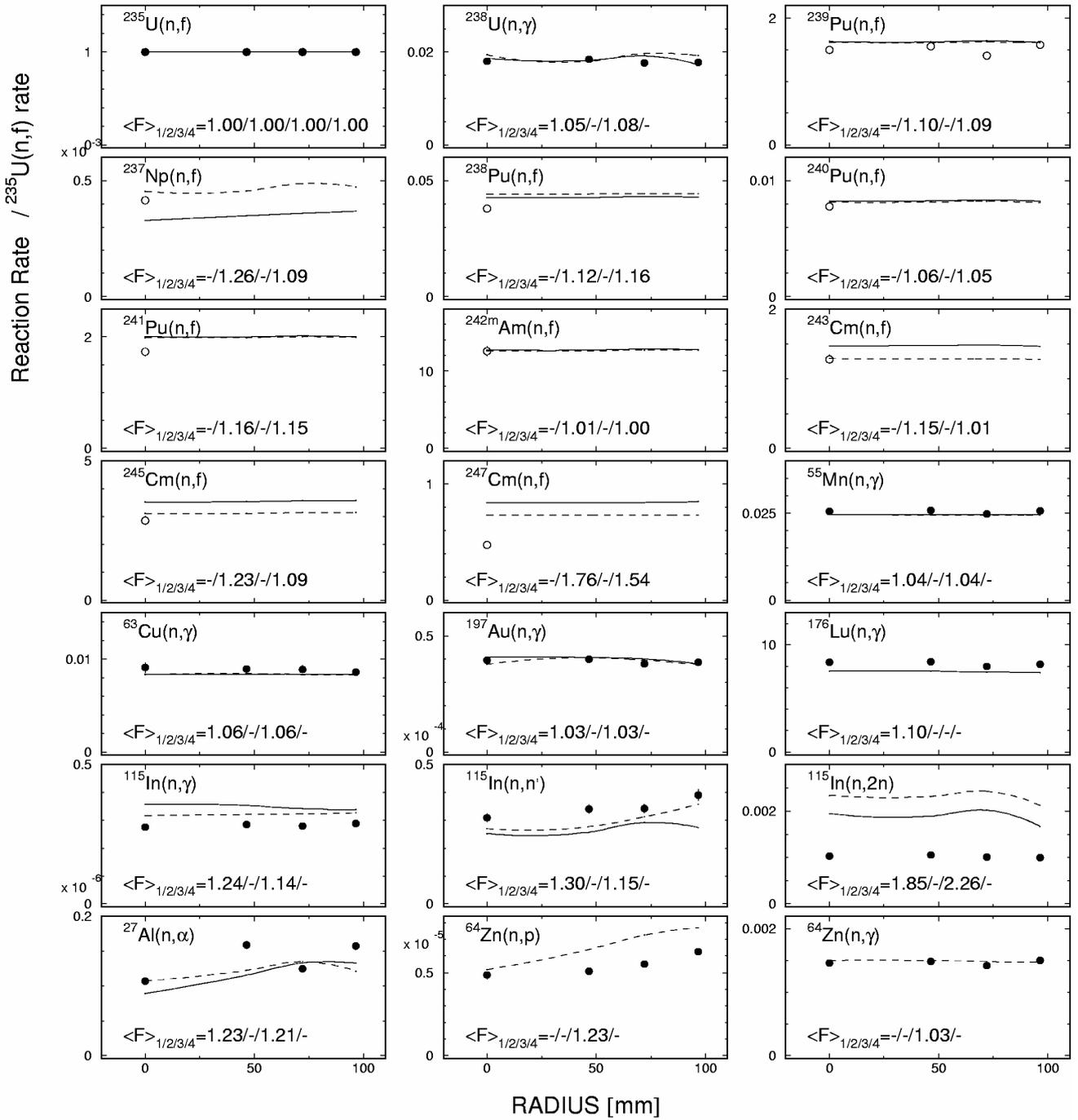

Fig. 4. The MAKET-measured and calculated reaction rates in the lattice with salt insert in central channel after normalizing to the $^{235}$U(n,f) reaction rate at the respective points. The black circles are the ITEP data. The light circles are the VNIIEF data. The dashed lines are the JENDL-3.2 –based calculations. The solid lines are the ENDF/B-IV–based calculations. $<F>_{1/2/3/4}$ are the experiment-calculation mean square deviation factors, where $<F>_1$ is the ITEP experiment- ENDF/B-VI calculation difference; $<F>_2$ is the VNIIEF experiment- ENDF/B-VI calculation difference; $<F>_3$ is the ITEP experiment- JENDL-3.2 calculation difference; $<F>_4$ is the VNIIEF experiment- JENDL-3.2 calculation difference. The gaps indicate lack of the appropriate experimental or calculated results.

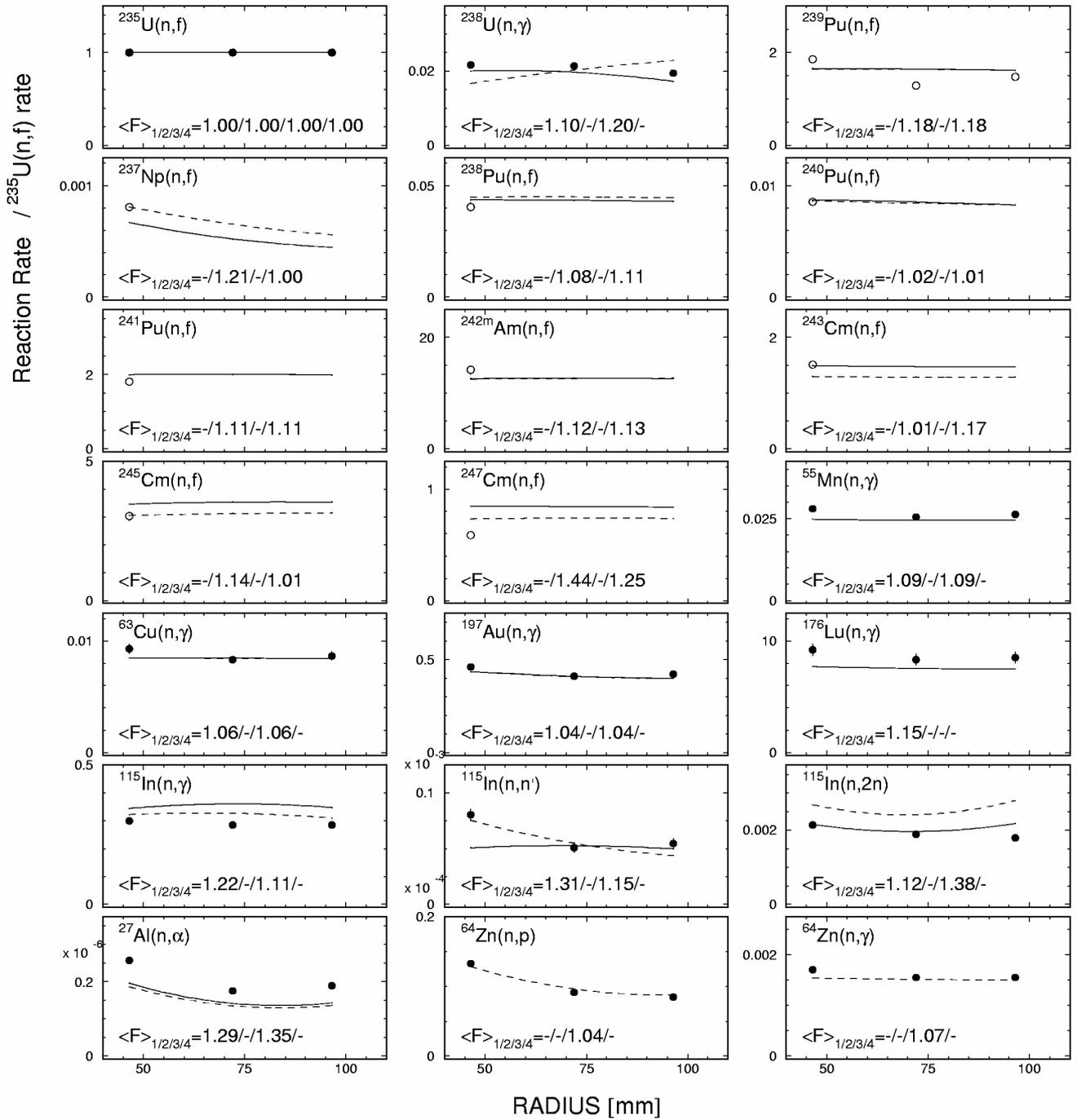

Fig. 5. The MAKET-measured and calculated reaction rates in the lattice with fuel elements in central channel after normalizing to the $^{235}$U(n,f) reaction rate at the respective points. The black circles are the ITEP data. The light circles are the VNIIEF data. The dashed lines are the JENDL-3.2 –based calculations. The solid lines are the ENDF/B-IV–based calculations. $<F>_{1/2/3/4}$ are the experiment-calculation mean square deviation factors, where $<F>_1$ is the ITEP experiment- ENDF/B-VI calculation difference; $<F>_2$ is the VNIIEF experiment- ENDF/B-VI calculation difference; $<F>_3$ is the ITEP experiment- JENDL-3.2 calculation difference; $<F>_4$ is the VNIIEF experiment- JENDL-3.2 calculation difference. The gaps indicate lack of the appropriate experimental or calculated results.

Table II. Mean squared factors of the simulated-experimental reaction rates deviations (absolute normalization).

| Reaction | ENDF/B-VI rev 7. | | | | JENDL-3.2 | | | |
|---|---|---|---|---|---|---|---|---|
| | 21-1-1-5(M2) | | 21-2 | | | | 21-1-1-5(M2) | |
| | ITEP | VNIIEF | | ITEP | VNIIEF | | ITEP | VNIIEF |
| $^{235}$U(n,f) | 1.15 | 1.17 | 1.06 | 1.25 | 1.14 | 1.16 | 1.06 | 1.25 |
| $^{238}$U(n,γ) | 1.13 | - | 1.17 | - | 1.09 | - | 1.21 | - |
| $^{239}$Pu(n,f) | - | 1.11 | - | 1.19 | - | 1.10 | - | 1.19 |
| $^{237}$Np(n,f) | - | 1.53 | - | 1.37 | - | 1.10 | - | 1.13 |
| $^{238}$Pu(n,f) | - | 1.08 | - | 1.06 | - | 1.04 | - | 1.03 |
| $^{240}$Pu(n,f) | - | 1.14 | - | 1.11 | - | 1.15 | - | 1.12 |
| $^{241}$Pu(n,f) | - | 1.05 | - | 1.03 | - | 1.05 | - | 1.03 |
| $^{242m}$Am(n,f) | - | 1.20 | - | 1.27 | - | 1.20 | - | 1.28 |
| $^{243}$Cm(n,f) | - | 1.05 | - | 1.15 | - | 1.20 | - | 1.33 |
| $^{245}$Cm(n,f) | - | 1.02 | - | 1.01 | - | 1.10 | - | 1.13 |
| $^{247}$Cm(n,f) | - | 1.45 | - | 1.27 | - | 1.28 | - | 1.10 |
| $^{55}$Mn(n,γ) | 1.18 | - | 1.15 | - | 1.18 | - | 1.15 | - |
| $^{63}$Cu(n,γ) | 1.22 | - | 1.10 | - | 1.21 | - | 1.11 | - |
| $^{197}$Au(n,γ) | 1.12 | - | 1.10 | - | 1.15 | - | 1.10 | - |
| $^{176}$Lu(n,γ) | 1.26 | - | 1.21 | - | - | - | - | - |
| $^{115}$In(n,γ) | 1.08 | - | 1.14 | - | 1.02 | - | 1.05 | - |
| $^{115}$In(n,n′) | 1.47 | - | 1.35 | - | 1.30 | - | 1.19 | - |
| $^{115}$In(n,2n)+ $^{113}$In(n,γ) | 1.58 | - | 1.07 | - | 1.95 | - | 1.28 | - |
| $^{27}$Al(n,α) | 1.37 | - | 1.37 | - | 1.31 | - | 1.43 | - |
| $^{64}$Zn(n,p) | - | - | - | - | 1.12 | - | 1.05 | - |
| $^{64}$Zn(n,γ) | - | - | - | - | 1.12 | - | 1.12 | - |

Table III. Mean squared factors of the simulated-experimental reaction rates deviations (relative normalization).

| Reaction | ENDF/BVI rev 7. | | | | JENDL-3.2 | | | |
| --- | --- | --- | --- | --- | --- | --- | --- | --- |
| | 21-1-1-5(M2) | | 21-2 | | 21-1-1-5(M2) | | 21-2 | |
| | ITEP | VNIIEF | ITEP | VNIIEF | ITEP | VNIIEF | ITEP | VNIIEF |
| $^{235}$U(n,f) | 1 | 1 | 1 | 1 | 1 | 1 | 1 | 1 |
| $^{238}$U(n,γ) | 1.05 | - | 1.10 | - | 1.08 | - | 1.20 | - |
| $^{239}$Pu(n,f) | - | 1.10 | - | 1.18 | - | 1.09 | - | 1.18 |
| $^{237}$Np(n,f) | - | 1.26 | - | 1.21 | - | 1.09 | - | 1.00 |
| $^{238}$Pu(n,f) | - | 1.12 | - | 1.08 | - | 1.16 | - | 1.11 |
| $^{240}$Pu(n,f) | - | 1.06 | - | 1.02 | - | 1.05 | - | 1.01 |
| $^{241}$Pu(n,f) | - | 1.16 | - | 1.11 | - | 1.15 | - | 1.11 |
| $^{242m}$Am(n,f) | - | 1.01 | - | 1.12 | - | 1.00 | - | 1.13 |
| $^{243}$Cm(n,f) | - | 1.15 | - | 1.01 | - | 1.01 | - | 1.17 |
| $^{245}$Cm(n,f) | - | 1.23 | - | 1.14 | - | 1.09 | - | 1.01 |
| $^{247}$Cm(n,f) | - | 1.76 | - | 1.44 | - | 1.54 | - | 1.25 |
| $^{55}$Mn(n,γ) | 1.04 | - | 1.09 | - | 1.04 | - | 1.09 | - |
| $^{63}$Cu(n,γ) | 1.06 | - | 1.06 | - | 1.06 | - | 1.06 | - |
| $^{197}$Au(n,γ) | 1.03 | - | 1.04 | - | 1.03 | - | 1.04 | - |
| $^{176}$Lu(n,γ) | 1.10 | - | 1.15 | - | - | - | - | - |
| $^{115}$In(n,γ) | 1.24 | - | 1.22 | - | 1.14 | - | 1.11 | - |
| $^{115}$In(n,n') | 1.30 | - | 1.31 | - | 1.15 | - | 1.15 | - |
| $^{115}$In(n,2n)+ $^{113}$In(n,γ) | 1.85 | - | 1.12 | - | 2.26 | - | 1.38 | - |
| $^{27}$Al(n,α) | 1.23 | - | 1.29 | - | 1.21 | - | 1.35 | - |
| $^{64}$Zn(n,p) | - | - | - | - | 1.23 | - | 1.04 | - |
| $^{64}$Zn(n,γ) | - | - | - | - | 1.03 | - | 1.07 | - |

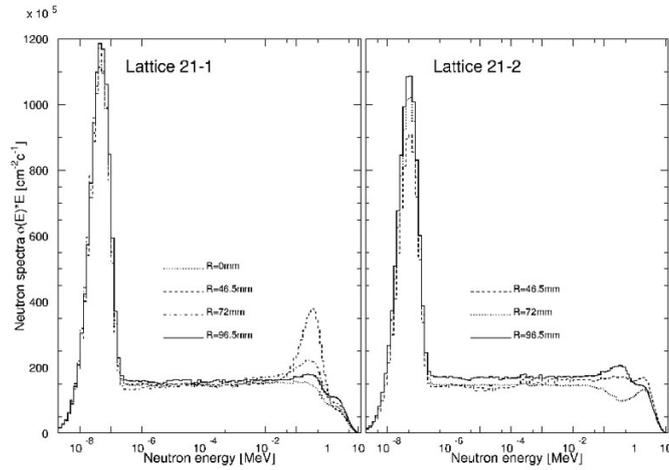

Fig 6. Neutron spectra in the SBM channels computed by theCASCADE code.

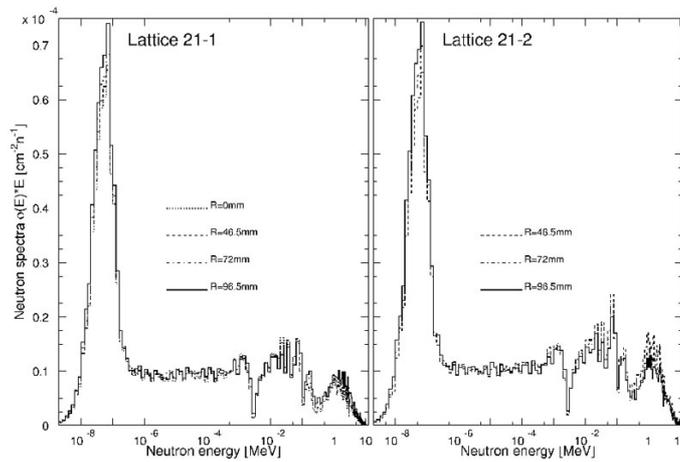

Fig 7. Neutron spectra in the SBM channels computed by the MCNP code.

Table IV. Preference in using the ENDF/B-VI and JENDL-3.2 libraries for two lattices in the absolute and relative normalizations

| Reaction | 21-1 | | | | 21-2 | | | |
| --- | --- | --- | --- | --- | --- | --- | --- | --- |
| | ENDF/BVI rev 7. | | JENDL-3.2 | | ENDF/BVI rev 7. | | JENDL-3.2 | |
| | Abs. Norm. | Rel. norm. | Abs. Norm. | Rel. norm. | Abs. Norm. | Rel. norm. | Abs. Norm. | Rel. norm. |
| $^{235}$U(n,f) | + | | + | | + | | + | |
| $^{238}$U(n,γ) | + | + | + | + | + | + | - | - |
| $^{239}$Pu(n,f) | + | + | + | + | + | + | + | + |
| $^{237}$Np(n,f) | - | - | + | + | - | - | + | + |
| $^{238}$Pu(n,f) | + | + | + | + | + | + | + | + |
| $^{240}$Pu(n,f) | + | + | + | + | + | + | + | + |
| $^{241}$Pu(n,f) | + | + | + | + | + | + | + | + |
| $^{242m}$Am(n,f) | + | + | + | + | + | + | + | + |
| $^{243}$Cm(n,f) | + | - | - | + | + | + | - | - |
| $^{245}$Cm(n,f) | + | - | - | + | + | - | - | + |
| $^{247}$Cm(n,f) | - | - | + | + | - | - | + | + |